\documentclass[amsmath,amssymb,nofootinbib,preprint]{revtex4}

\usepackage{latexsym,comment}
\usepackage{amssymb}
\usepackage{amsfonts}
\usepackage{amsmath}
\usepackage[dvips]{graphicx}
\usepackage{bm}

\def\ber{\begin{eqnarray}}
\def\eer{\end{eqnarray}}
\def\beq{\begin{equation}}
\def\eeq{\end{equation}}

\def\eg{\`e}

\def\ea{\'e}
\DeclareSymbolFont{extraup}{U}{zavm}{m}{n}
\DeclareMathSymbol{\varheart}{\mathalpha}{extraup}{86}
\DeclareMathSymbol{\vardiamond}{\mathalpha}{extraup}{87}

\begin{document}

\title{Sagnac effect and pure geometry}

\author{Angelo Tartaglia}
\email[Corresponding Author:\ ]{angelo.tartaglia@polito.it}
 \affiliation{DISAT, Politecnico di Torino, Corso Duca degli Abruzzi 24, Torino, Italy\\
 INFN, Sezione di Torino, Via Pietro Giuria 1, Torino, Italy}
\author{Matteo Luca Ruggiero}
 \affiliation{DISAT, Politecnico di Torino, Corso Duca degli Abruzzi 24, Torino, Italy\\
 INFN, Sezione di Torino, Via Pietro Giuria 1, Torino, Italy}

\date{\today}

\begin{abstract}
We show that the Sagnac effect is not necessarily due to the presence of a rotating observer, but rather to the closed path of light in space and an even inertial relative motion between the observer and the physical device forcing light to move along a closed path.
\end{abstract}

\maketitle

\section{Introduction}

The Sagnac effect takes its name from Georges Marc Marie Sagnac, a French physicist who, in 1913, partially interpreting previous experimental evidence, showed that the time of flight of light emitted by a source on a rotating platform and sent back by means of mirrors along a path closed in space was asymmetric. The asymmetry between the forward (in the sense of rotation) and backward beams, in terms of times of flight difference, is proportional to the angular velocity of the platform and appears as a phase difference measured by an interferometer. Sagnac interpreted this result as an evidence \textit{against} Einstein's relativity and in favour of a static luminiferous ether.\cite{sagnac1}\cite{sagnac2} It was however immediately and easily shown that the Sagnac effect is indeed a fully relativistic effect.

Today the Sagnac effect is relevant for various applications, all related to the measurement of rotation rates. On the commercial side we have gyrolasers, developed since the 70's of the last century, used on planes, ships, submarines, missile guidance systems... The name 'gyrolaser' is motivated by the fact that the devices replace old mechanical gyroscopes and use a laser to produce the counter-rotating light beams. Instead of exploiting the interference between the two opposed light rays, they measure the beat frequency of the standing wave resulting from the superposition of two waves propagating in opposite directions; that frequency is again proportional to the absolute angular velocity of the apparatus (a thorough  description of gyrolaser devices can be found in the review paper \cite{chow85}).

The most sensitive ring lasers (a.k.a. gyrolasers) sense the diurnal rotation of the earth, the wobbles of the terrestrial axis, the tiny rotations of the laboratory hosting the instrument, due to elastic deformations of the ground caused by liquid and solid tides and also by various surface phenomena. These rings are used for high accuracy geophysical measurements.\cite{ulli}

The extremely high sensitivity obtained in most recent ring lasers is such that now the Sagnac effect is  becoming important also for fundamental physics: the possibility has been  suggested  of using large ring lasers to detect the Lense-Thirring effect on Earth \cite{StedLT} and an experiment, under the acronym GINGER, is now in the phase of development and preliminary testing at the National Laboratories of the Italian INFN at the Gran Sasso site.\cite{ginger11}\cite{ginger12}
The Sagnac effect plays a role also in the elaboration of the position of a receiver obtained from the GPS: a Sagnac correction is indeed required because a rotating clock (onboard a satellite or on the surface of the earth) suffers a loss of synchrony with itself at each turn. The relevance of the relativistic effects on clocks carried around, either in the same or in the opposite sense with respect to the rotation of the earth, was experimentally verified by Hafele and Keating.\cite{HK} \cite{HK1} \cite{RSch}

The Sagnac effect has been deeply investigated, soon after the realization of Sagnac's experiments, as a disproval of the theory of relativity, and also more recently, in connection with the issue of synchronization in rotating reference frames (see for instance \cite{abs} and the monograph \cite{RRF}). All this is well known, but it is also accompanied by a common feeling that the Sagnac effect is essentially due to the presence of a rotating observer or rotating device: in short, to non-inertial motion. Examples of papers elaborating on various aspects of the Sagnac effect, without discussing the issue of the necessity of rotation are \cite{AZS} \cite{GBab} \cite{KKa}.

In  reality, however, non-inertial motion is not necessary. Indeed, a few years ago some experimental papers have been published  which  claim that even if no rotation is present, a ``generalized Sagnac effect'' arises in a uniformly moving fiber.\cite{WZYL,prlbis} In the present paper we aim at addressing the issue, that is we want to show that the real ingredients of the Sagnac effect are two: a) a closed circuit followed by light in opposite directions; b) a relative (even inertial) motion of the emitter/receiver with respect to the physical apparatus supporting the closed path of light. We believe that this approach could be useful in teaching the foundations of relativity, since the calculations involved are simple, and they allow a deep insight into the physics of the problem, that ultimately is connected with  the relativity of simultaneity.

\section{The Sagnac effect} \label{section:sect1}

The physical principles of the Sagnac effect are explained in full details in the well known paper by Post,\cite{post67} and they can be summarized as follows.  The Sagnac effect may easily be described in classical terms if one assumes that the speed of light is $c$ with respect to a static ether. Considering the rotating platform mentioned in the Introduction you see that it will take longer for light to reach again the emission point on the rim of the platform just because, meanwhile, the receiver will have moved forward by a distance $\Delta l_+=vt_+$ where $t_+$ is the total time of flight and $v$ is the velocity of the emitter with respect to the ether; the geometric length of the path is $l$. Hence the time of flight is read from the following relation $\displaystyle \frac{l + \Delta l_{+}}{c}= \frac{\Delta l_{+}}{v}$. In the opposite sense the receiver will move towards the returning light so that the path will be shorter by $\Delta l_-=vt_-$, whence $\displaystyle \frac{l - \Delta l_{-}}{c}= \frac{\Delta l_{-}}{v}$. Solving for $t_+$ and $t_-$ gives the time of flight difference:

\begin{equation}
\Delta t=t_+-t_-=\frac{2lv}{c^2-v^2}
\label{classic}
\end{equation}

An equivalent deduction can be made adopting the viewpoint of the rotating observer. In his/her reference frame light will be expected to have speed $c-v$ in one sense (forward) and $c+v$ in the other; now the path is the same $l$ for both. The times of flight are again different and a trivial calculation brings back again to formula (\ref{classic}). Add that in the example we have $v=\omega R$ with $\omega$ representing the angular velocity of the platform and $R$ the distance of the observer from the rotation axis; simplify the geometry assuming that the path of light is a circle\footnote{This would have been practically impossible in Sagnac's time, but is feasible today with modern optical fibers, as it is done in commercial gyrolasers.}  at radius $R$ so that $l=2\pi R$; consider that in all practical cases we have $v<<c$, and Eq. (\ref{classic}) will be transformed into:

\begin{equation}
\Delta t\simeq 4\frac{\mathbf{A \cdot \bm{\omega}}}{c^2}
\label{sagnac0}
\end{equation}

Here $\bf{A}$ is the area enclosed in the path of the light beams, represented as a vector, whose magnitude is $\pi R^{2}$, perpendicular to the plane containing the trajectory \footnote{Actually the planarity is not needed.}; $\bm{\omega}$ is also a vector and the dot product projects the area in a plane orthogonal to the rotation axis. Equation (\ref{sagnac0}) is the famous Sagnac formula, where the $\simeq$ symbol is currently replaced by a plain $=$.

\begin{figure}[top]
\begin{center}
\includegraphics[scale=.30]{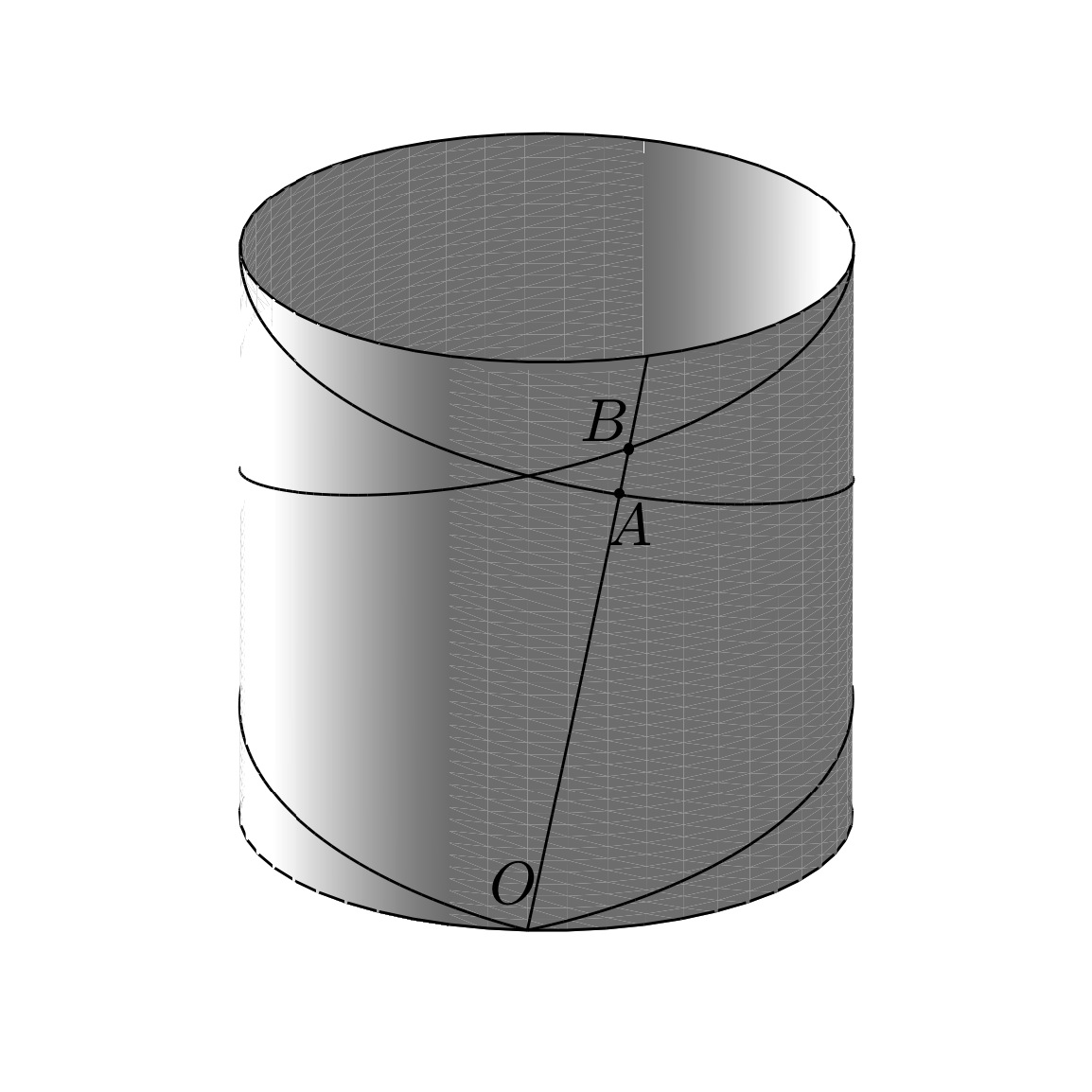}
\end{center}
\caption{Spacetime diagram of the Sagnac effect. $OAB$ is the
world line of the rotating observer. The line is a helix drawn on
a three-dimensional cylinder. Along the vertical axis inertial time is measured.
The other two lines (helices) represent two light
rays. $A$ and $B$ are the events where the observer is reached by
the counter-, respectively, co-rotating beams.}
\label{fig:fig1}
\end{figure}

As said, however, the Sagnac effect does not need any ether at all. Special (and general) relativity is essentially geometry so we may use a geometric approach in order to work out what happens.  In the simplified configuration of the classical example the situation may be represented in a spacetime graph like the one in Fig. (\ref{fig:fig1}).

Everything is drawn on the surface of a cylinder and the viewpoint is that of an inertial observer at rest with respect to the axis of the rotating platform.\cite{grgmio} The vertical axis of the figure is \textit{not} the rotation axis of the turntable; it is the coordinate time axis. The worldline of the rotating observer is a helix and the two light beams are represented by two opposite helices as well, which, after one turn, intersect the worldline of the observer at two different events ($A$ and $B$, in the figure). The interval between $A$ and $B$ is the proper time difference measured by the observer between the arrival times of the two light rays.

\begin{figure}
\begin{center}
\includegraphics[scale=.90]{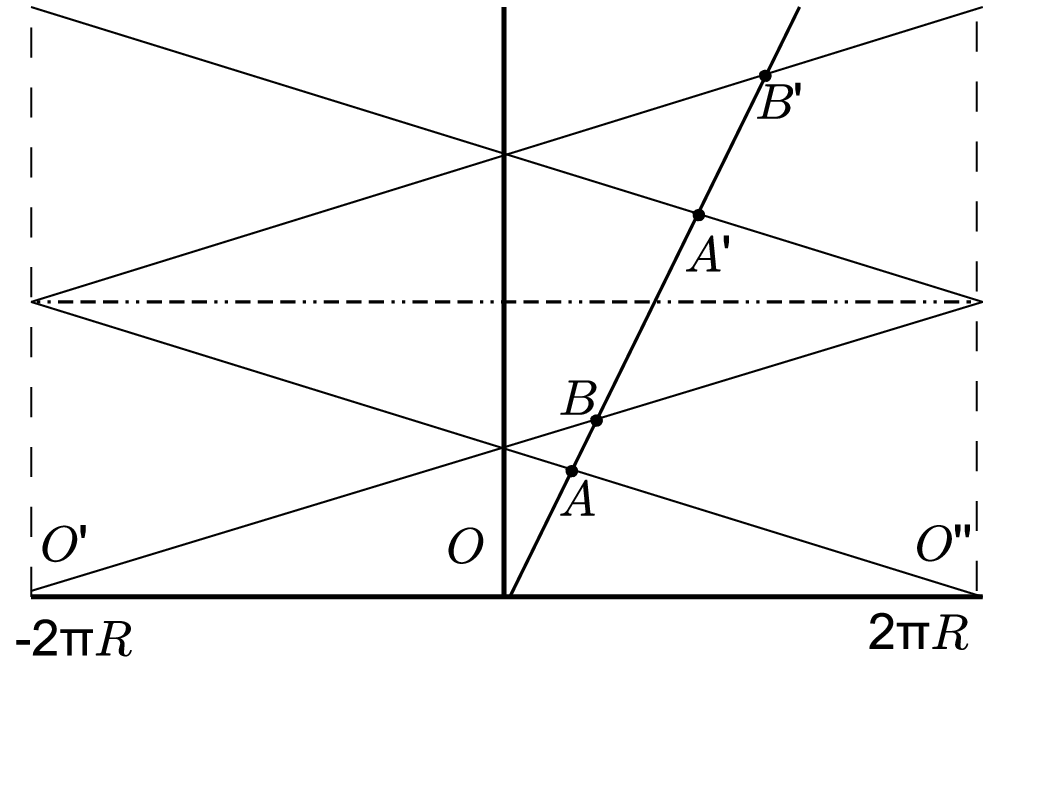}
\end{center}
\caption{Same as Fig. \ref{fig:fig1}. The cylinder has been cut along a generatrix passing through the rotating observer at time 0,
and opened. For convenience, in order to make the picture more compact, two replicas of the opened cylinder are shown side by side: one enrolled to the right, the other to the left. Points $O'$ and $O''$ coincide with $O$. Four windings are shown. The vertical straight line is the world
line of an inertial observer at rest with the axis of the disk. $AB$ is the
Sagnac effect expressed in terms of proper time of the rotating observer.}
\label{fig:fig2}
\end{figure}

A cylinder is indeed a flat surface, so we may have a simpler view cutting the cylinder along one generatrix, then opening it in a flat spacetime strip like in Fig. \ref{fig:fig2}. Now helices become straight lines and the geometry is evident and simple; the only trick to remember is that the right border of the strip is identified with the left border, so that lines reaching one border reappear at the corresponding point on the other.

The properties of straight lines and triangles are the usual ones, recalling that this is not Euclidean space, but Minkowski spacetime so that the squared length of the hypotenuse is now a squared proper time multiplied by $c^2$ and is given by the \textit{difference} between the squared coordinate time span times $c^2$ and the square of the travelled distance. Pure geometry \cite{grgmio} leads again, at the lowest approximation order, to Eq. (\ref{sagnac0}).

If one prefers to proceed analytically the starting point can be the line element of Minkowski spacetime written using cylindrical coordinates in space: $ds^2 = c^2dt^2-dr^2-r^2d\phi^2-dz^2$. Introducing the constraints $r = R = constant$ and $z = constant$, corresponding to all physics taking place on the rim of a disk perpendicular to the $z$ axis, the problem is reduced to two effective dimensions and the line element becomes:

\begin{equation}
ds^2=c^2dt_0^2-R^2d\phi_0^2
\label{line}
\end{equation}

The label $_0$ means that the coordinates are those of the inertial observer.

To account for the rotation of the platform it is convenient to introduce axes rotating together with the platform so that the angular coordinate becomes $\phi = \phi_0-\omega t_0$. Then we need a Lorentz transformation between the inertial observer at rest with the axis of the platform and the inertial frame instantaneously comoving with the observer on the rim of the turntable. In Special Relativity the Lorentz transformations leave the line element unchanged, so that in the rotating reference frame it becomes:

\begin{equation}
ds^2=(c^2-\omega^2R^2)dt^2-R^2d\phi^2-2R^2 \omega d\phi dt
\label{line1}
\end{equation}

On writing the previous expression in the form
\beq
ds^{2}=g_{00}dt^{2}+g_{\phi\phi}d\phi^2+2g_{0\phi}dtd\phi \label{eq:metricastazionaria}
\eeq
and considering that for light we have $ds=0$, it is possible to work out $dt$:
\beq
dt=\frac{-g_{0\phi} d\phi \pm \sqrt{g^{2}_{0\phi}d\phi^{2}-g_{\phi\phi}g_{00}d\phi^2}}{g_{00}}
\label{eq:1rev}
\eeq
We are interested in solutions located in the future, so we choose $dt > 0$, i.e. the $+$ sign in the formula. Under the simplified assumption that light moves along a circumference, if we integrate in the two opposite directions (counterclockwise, $d\phi>0$; clockwise, $d\phi<0$) from the emission to the absorption events, we get the expression for the co-rotating  ($t_{+}$) and counter-rotating  ($t_{-}$) times of flight, and the difference between them turns out to be
\beq
\Delta t= t_{+}-t_{-} = -2 \oint_{\ell} \frac{g_{0\phi}}{g_{00}}  R d\phi
\eeq
where $\ell$ is the circumference of radius $R$ \footnote{In the most general case $\ell$ is the integration path and, using arbitrary coordinates, the formula would be $\Delta t= -2 \oint_{\ell} \frac{g_{0i}}{g_{00}}  dx^i$ (see e.g. \cite{shorty}).}. Moreover, the observer   measures  the proper-time difference:
\beq
\Delta \tau=-2 \sqrt{g_{00}}  \oint_{\ell} \frac{g_{0\phi}}{g_{00}}  R d\phi \label{eq:deltataulocal1}
\eeq
On substituting the expressions (\ref{line1}) of the metric coefficients, we get the following expressions for the coordinate and proper times of flight difference
\begin{eqnarray}
\Delta t &=& 4\frac{\pi R^{2} \omega}{c^2(1-\omega^2 R^2/c^2)}  \nonumber \\
\Delta \tau &=& 4\frac{\pi R^{2} \omega}{c^2\sqrt{1-\omega^2 R^2/c^2}}  \nonumber
\end{eqnarray}

At the lowest approximation order in $\omega R/c$ the two coincide, and they are in agreement with Eq. (\ref{sagnac0}), with $\mathbf{A}=\pi R^{2}\hat{\mathbf u}_{z}$. The unit vector $\hat{\mathbf u}_{z}$ points along the $z$ direction.

\section{Sagnac without rotation}

\begin{figure}[th]
\begin{center}
\includegraphics[width=5cm,height=5cm]{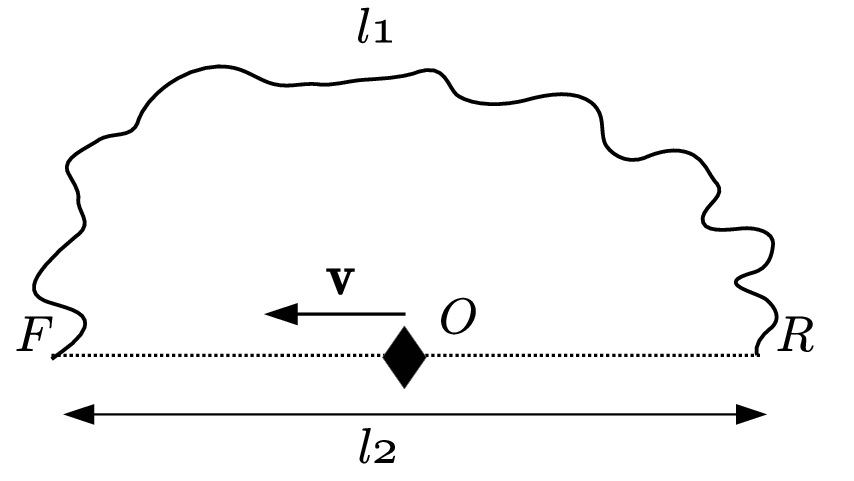}
\end{center}
\caption{$O$ represents a source/receiver of light. Light is sent both
forward and backward (dashed lines). The irregular line joining $F$ with $R$
represents an optical fiber of length $l_{1}$. $O$ is moving with the
velocity $v$ with respect to the fiber. The total length of the path is of course $L=l_{1}+l_{2}$}.
\label{fig:fig3}
\end{figure}

Despite the widespread tendency to ascribe the Sagnac effect to rotating systems it is easy to show that rotation is not essential. Consider a source/receiver of light in inertial motion at the speed $v$ with respect to a set of mirrors rigidly fastened to one another or to an optical fiber so that they guide the light beams emitted by the moving source along a closed path in space. The situation is schematically shown in Fig. \ref{fig:fig3} where an optical fiber is envisaged.

Repeating the simple reasoning made at the beginning of sec. \ref{section:sect1} we easily arrive to the same result written in formula (\ref{classic}). The only additional clarification is that Eq. (\ref{classic}) is expressed from the view point of an observer at rest with the fiber (or array of mirrors); in the proper time of the moving emitter/receiver it would rather be (we set $\displaystyle \beta=\frac{v}{c}$):

\begin{equation}
\Delta \tau =2\frac{Lv}{c^{2}\sqrt{1-\beta^{2}}}\sim 2\frac{Lv}{%
c^{2}}
\label{sagnac4}
\end{equation}

No rotation, no acceleration, no enclosed area appear. What matters is just the existence of relative (even inertial) motion and of a closed path in space. Of course, with reference to Fig. \ref{fig:fig3}, $%
\bf{v}$ is parallel to the light rays in $RF$.

The effect we have just described and the related formula (\ref{sagnac4}) have been verified by experiment by Wang, Zheng, Yao, and Langley (WZYL).\cite{WZYL,prlbis}

Using the geometrical approach the result is rather obvious. In spacetime
the world lines are drawn again on a cylinder. The only difference is that
on Fig. \ref{fig:fig1} the cylinder had a circular cross section, now the cross section
is rather irregular. In any case the cylinder is a flat bidimensional
Minkowski surface and, when opening it in a plane, the graph is exactly like in
Fig. \ref{fig:fig2}.

Formula (\ref{classic}) appears also in Ref. \cite{ashby} where it is related to the relativity of simultaneity typical of special relativity.

Another aspect to clarify is related to the form (\ref{sagnac0}) of the effect. The accidental presence of the area $A$ suggested to some people \cite{stedman} \cite{EGH} an interpretation of the Sagnac effect (at least for matter waves) as a sort of Aharonov-Bohm effect: however, this is just an analogy, which holds at lowest relativistic order only (see e.g. \cite{shorty}).

\section{A simple example}\label{example}

\begin{figure}[here]
\begin{center}
\includegraphics[scale=.50]{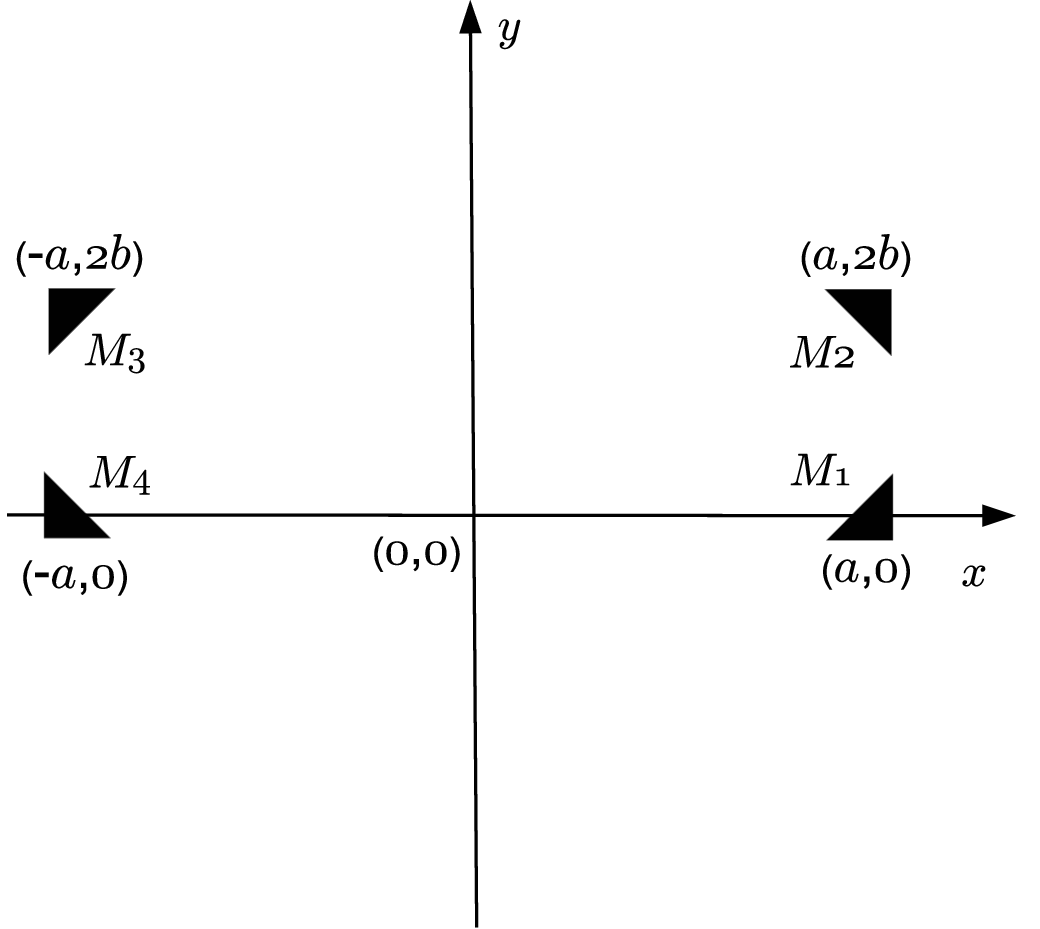}
\end{center}
\caption{The four mirrors $M_{1}$, $M_{2}$, $M_{3}$, $M_{4}$ that form the interferometer are at rest in the laboratory frame.}
\label{fig:setup}
\end{figure}

\begin{figure}[here]
\begin{center}
\includegraphics[scale=.50]{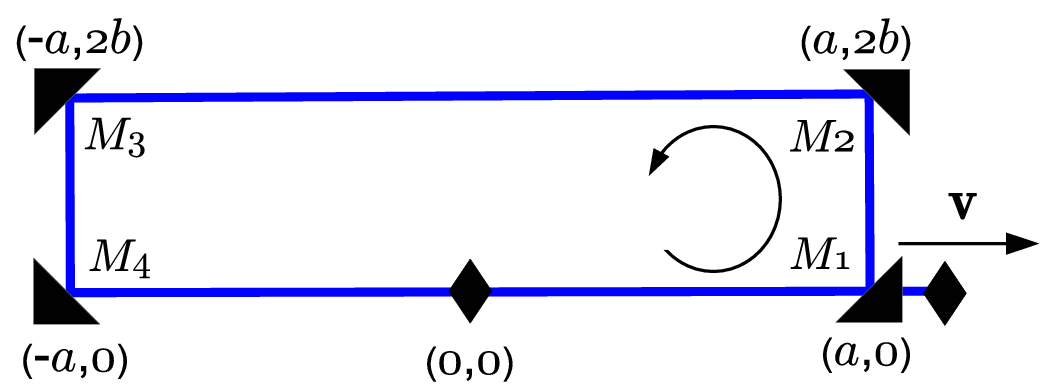}
\end{center}
\caption{CCW rays path, as seen in the laboratory; the symbol $\vardiamond$ denotes the position of the observer when he sends the signal and when he is reached by the signal. To draw the figure, we chose $\beta=20/101$, so that $\gamma=101/99$, and $b=a/4$.}
\label{fig:restccw}
\end{figure}

\begin{figure}[here]
\begin{center}
\includegraphics[scale=.50]{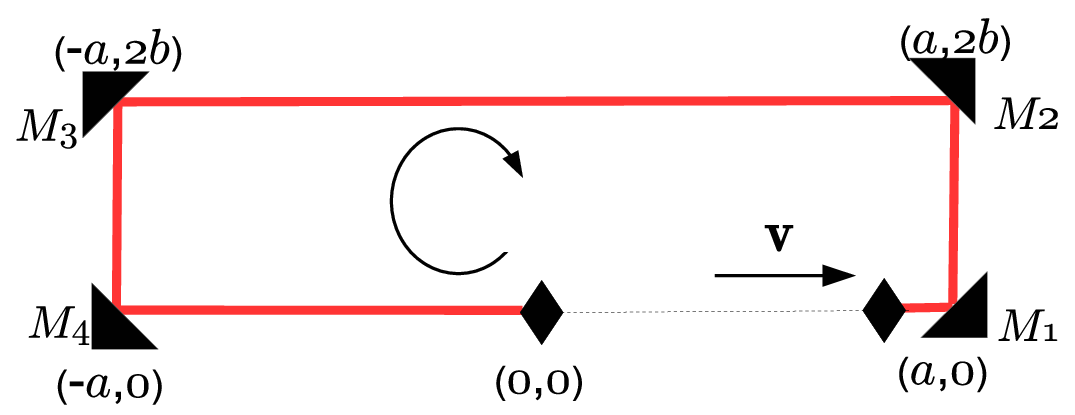}
\end{center}
\caption{CW rays path, as seen in the laboratory; the symbol  $\vardiamond$  denotes the position of the observer when he sends the signal and when he is reached by the signal. To draw the figure, we chose $\beta=20/101$, so that $\gamma=101/99$, and $b=a/4$.}
\label{fig:restcw}
\end{figure}

In order to get a deeper insight, let us consider a simple example where the calculation can be carried out in full details.
Our interferometer is at rest in the laboratory and is made of four mirrors $M_{1}$, $M_{2}$, $M_{3}$, $M_{4}$, at the corners of a rectangle: if we use Cartesian coordinates, their positions are given in Fig. \ref{fig:setup}, and the interferometer path length is $L=4a+4b$.  An observer, moving at speed $v$ to the right  in the laboratory frame, sends two counter propagating light signals when he is at the origin. The paths of the counterclockwise (CCW) and clockwise rays (CW) are depicted in Fig.s \ref{fig:restccw} and \ref{fig:restcw}, respectively, as seen in the laboratory frame. Note that in the laboratory frame, the CW path \textit{is not closed}, and the CCW path \textit{overlaps} with itself. What we want to show, by the following calculation, is that  \textit{both paths are closed} in the frame moving with the observer.

In terms of Cartesian spacetime coordinates $(ct,x,y)$ in the laboratory frame, the events sequence along the CCW path is
\begin{itemize}
\item $e_{0}$: $(0,0,0)$ (emission)
\item $e_{1}$: $(a,a,0)$ (reflection by $M1$)
\item $e_{2}$: $(a+2b,a,2b)$ (reflection by $M2$)
\item $e_{3}$: $(3a+2b,-a,2b)$ (reflection by $M3$)
\item $e_{4}$: $(3a+4b,-a,0)$ (reflection by $M4$)
\item $e_{5}$: $(ct_{1},v t_{1},0)=\left(\frac{4\left(a+b\right)}{1-\beta}, \frac{4\left(a+b\right)}{1-\beta}\beta,0 \right)$ (reception)
\end{itemize}
In particular, the total propagation time is obtained from $ct_{1}=v t_{1}+4a+4b$.

As for the CW path, we have the following events:
\begin{itemize}
\item $e_{0}$: $(0,0,0)$ (emission)
\item $E_{1}$: $(a,-a,0)$ (reflection by $M4$)
\item $E_{2}$: $(a+2b,-a,2b)$ (reflection by $M3$)
\item $E_{3}$: $(3a+2b,a,2b)$ (reflection by $M2$)
\item $E_{4}$: $(3a+4b,a,0)$ (reflection by $M1$)
\item $E_{5}$: $(ct_{2},v t_{2},0)=\left(\frac{4\left(a+b\right)}{1+\beta}, \frac{4\left(a+b\right)}{1+\beta}\beta,0 \right)$ (reception)
\end{itemize}
Again, the total propagation time is obtained from $ct_{2}=4a+4b-v t_{2}$. As expected, the CCW is longer than the CW one, since $t_{1}>t_{2}$, so the two signals reach the observer at different times.

\begin{figure}[here]
\begin{center}
\includegraphics[scale=.50]{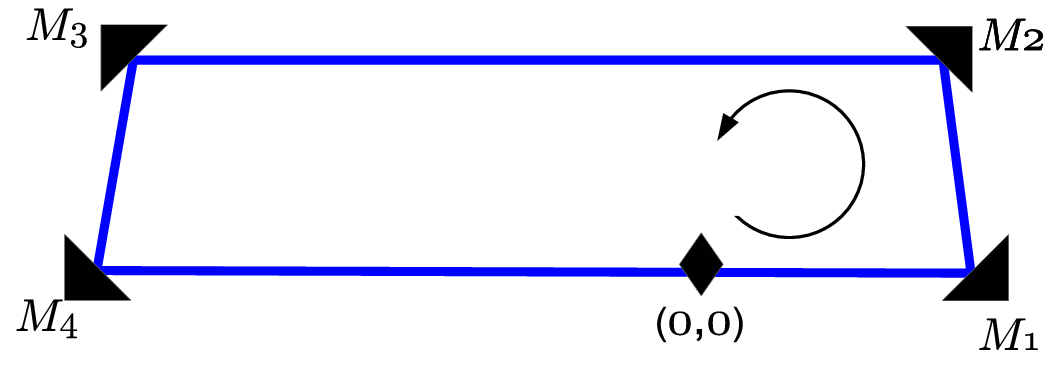}
\end{center}
\caption{CCW rays path, as seen in the observer's frame: both emission and reception take place at the origin $(0,0)$. To draw the figure, we chose $\beta=20/101$, so that $\gamma=101/99$, and $b=a/4$.}
\label{fig:movccw}
\end{figure}

\begin{figure}[here]
\begin{center}
\includegraphics[scale=.50]{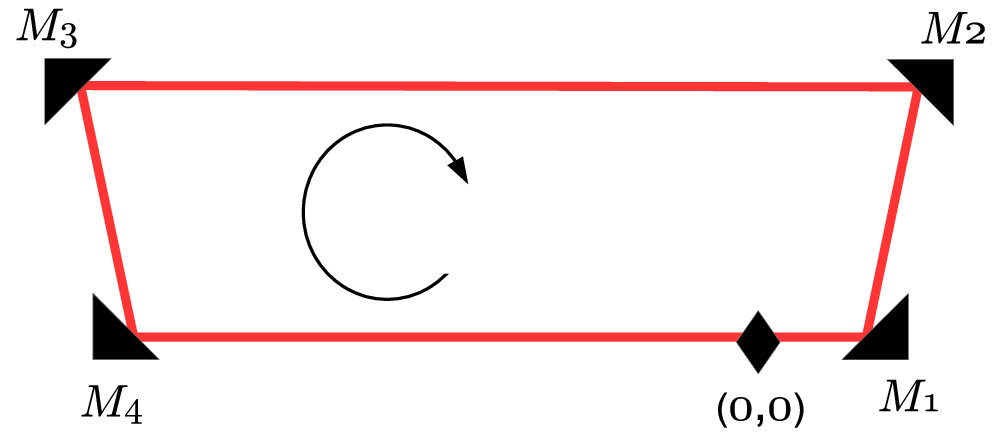}
\end{center}
\caption{CW rays path, as seen in the observer's frame: both emission and reception take place at the origin $(0,0)$. To draw the figure, we chose $\beta=20/101$, so that $\gamma=101/99$, and $b=a/4$.}
\label{fig:movcw}
\end{figure}

To obtain the coordinates of these events in the frame co-moving with the observer, we use the Lorentz transformations:
\beq
ct'=\gamma \left (ct- \beta x\right), \quad x'=\gamma \left (x- \beta ct\right), \quad y'=y \label{eq:lorentz}
\eeq
where $\displaystyle \gamma=\left(1-\beta^{2}\right)^{-1/2}$. As a consequence, the Cartesian  coordinates $(ct',x',y')$ of the events for the CCW path are
\begin{itemize}
\item $e_{0}$: $(0,0,0)$ (emission)
\item $e_{1}$: $\left(\gamma\left(1-\beta \right)a,\gamma\left(1-\beta \right)a,0 \right)$ (reflection by $M1$)
\item $e_{2}$: $\left(\gamma\left(1-\beta \right)a+2\gamma b,\gamma\left(1-\beta \right)a-2\gamma \beta b, 2b \right)$ (reflection by $M2$)
\item $e_{3}$: $\left(\gamma\left(3+\beta \right)a+2\gamma b,-\gamma\left(1+3\beta \right)a-2\gamma \beta b, 2b \right)$ (reflection by $M3$)
\item $e_{4}$: $\left(\gamma\left(3+\beta \right)a+4\gamma b,-\gamma\left(1+3\beta \right)a-4\gamma \beta b,0 \right)$ (reflection by $M4$)
\item $e_{5}$: $\left(4\gamma \left(1+\beta \right)(a+b),0,0 \right)$ (reception)
\end{itemize}
For the CW path we have the following coordinates:
\begin{itemize}
\item $e_{0}$: $(0,0,0)$ (emission)
\item $E_{1}$: $\left(\gamma\left(1+\beta \right)a,-\gamma\left(1+\beta \right)a, 0 \right)$ (reflection by $M4$)
\item $E_{2}$: $\left(\gamma\left(1+\beta \right)a+2\gamma b,-\gamma\left(1+\beta \right)a-2\gamma \beta b, 2b \right)$ (reflection by $M3$)
\item $E_{3}$: $\left(\gamma\left(3-\beta \right)a+2\gamma b,\gamma\left(1-3\beta \right)a-2\gamma \beta b, 2b \right)$ (reflection by $M2$)
\item $E_{4}$: $\left(\gamma\left(3-\beta \right)a+4\gamma b,\gamma\left(1-3\beta \right)a-4\gamma \beta b, 0 \right)$ (reflection by $M1$)
\item $E_{5}$: $\left(4\gamma \left(1-\beta \right)(a+b),0,0 \right)$ (reception)
\end{itemize}
These collections of events show that \textit{in the moving frame the two paths are closed}.\\ In particular, CCW path is indeed a polygon whose vertices have the following $(x',y')$ coordinates:  $\displaystyle \left( 0,0 \right) $, $\displaystyle \left( \gamma\left(1-\beta \right)a, 0 \right) $, $\displaystyle \left( \gamma\left(1-\beta \right)a-2\gamma \beta b, 2b \right) $, $\displaystyle \left( -\gamma\left(1+3\beta \right)a-2\gamma \beta b, 2b \right) $, $\displaystyle \left( -\gamma\left(1+3\beta \right)a-4\gamma \beta b,0 \right) $. The  shape of this polygon is depicted in Fig. \ref{fig:movccw} and its length is $\ell_{1}=4 \gamma \left(1+\beta \right)(a+b)$; since the traversal time is read from event $e_{5}$, $t_{1}'=4\gamma \left(1+\beta \right)\frac{(a+b)}{c}$, we see that the average traversal speed is $c$. \\
On the other hand, the CW path is a polygon whose vertices have the following
coordinates: $\displaystyle \left(0,0 \right) $, $\displaystyle \left(-\gamma\left(1+\beta \right)a, 0 \right) $, $\displaystyle \left(-\gamma\left(1+\beta \right)a-2\gamma \beta b, 2b \right) $, $\displaystyle \left(\gamma\left(1-3\beta \right)a-2\gamma \beta b, 2b \right) $, $\displaystyle \left( \gamma\left(1-\beta \right)a-4\gamma \beta b, 0 \right) $. The shape of this polygon is depicted in Fig. \ref{fig:movcw}, and its length is $\ell_{2}=4 \gamma \left(1-\beta \right)(a+b)$; again, by reading the traversal time $t_{2}'=4\gamma \left(1-\beta \right)\frac{(a+b)}{c}$ from $E_{5}$, we see that the average traversal speed is $c$.

We remark that we can easily explain the tilt of the lateral edges in Fig.s \ref{fig:movccw} and \ref{fig:movcw} in terms of the motion of the observer: indeed, when the light ray moves upward at right angles to the x axis in the laboratory frame, it has to move upward and \textit{left} in the observer's frame; similarly, when it moves downward, it has to move downward and \textit{left} in the observer's frame.\\

The moving observer measures the proper time difference between the propagation of the CCW and CW rays
\beq
\Delta \tau= t'_{1}-t'_{2}= 8 \gamma  \frac{(a+b)}{c} \beta \simeq 2  \frac{Lv}{c^{2}} \label{eq:exdeltatau}
\eeq
This result is in agreement with the general formula (\ref{sagnac4}). Moreover, this calculation allows to understand the origin of the time difference arising in the moving frame: the two light paths, which are congruent except for their end pieces in the laboratory frame, are no longer congruent in the observer's moving frame. Consequently, since they have different lengths, light must take different time to propagate through them.

This example shows how to explain the Sagnac-effect in the case of an inertial observer: indeed, a similar approach can be used to do the same thing, for an accelerated one. However, while in our case the observer's space is Euclidean and everything is unambiguous,  care is needed in defining the spatial geometry for an arbitrary accelerated observer: spatial geometry can be defined in terms of congruence of world lines (see e.g. \cite{relspace}), so it is not simply related to \textit{one} observer, rather it depends on a \textit{collection} of observers.

Things are simple in the case of the uniformly rotating disk, where the ordinary Sagnac effect usually is explained: a natural choice of observers that enables to define the spatial geometry is given by the observers sitting at fixed points on the disk (see again \cite{relspace} and \cite{kassner1}). In this case, both the co-rotating and the counter-rotating  rays propagate along the \textit{same curve}. Then, the Sagnac effect can be explained in terms of different speeds of light in the co-rotating and counter-rotating directions, if clocks along the rim of the disk are Einstein-synchronized, or in terms of a time gap, if a different synchronization procedure is used (see e.g. \cite{RRS} and \cite{KKa}).

\section{Possible applications}

Not considering the conceptual aspects, the phenomenon we have interpreted and WZYL have measured prompts possible interesting applications, as   suggested in \cite{prlbis}.

As we see from Eq. (\ref{sagnac4}) the time of flight difference for light is proportional to the relative velocity of the source with respect to the fiber or the mirrors. Now consider a ring laser where the active segment is not rigidly connected to the rest of the annular cavity; imagine, for instance, that the active cavity is attached to the rest via an elastic support allowing for relative vibrations. What is usually measured in a ring laser is a beat frequency of two counter-propagating beams. The beat frequency is easily obtained from Eq. (\ref{sagnac4}) and is:

\begin{equation}
\nu_b = \frac{v}{\lambda}
\label{battimento}
\end{equation}

where $\lambda$ is the wavelength of the laser.
We have a simple speedometer for instantaneous velocities where 1 m/s corresponds roughly to beat frequencies of the order of MHz; but in the obvious case that the active element is subject to an acceleration, a simple differentiation of the output converts the device into an accelerometer.

Of course practical analyses on the stability of the laser modes should be made, but the idea is appealing.

\section{Conclusion}

We have shown that the Sagnac effect is due to the closure of the path followed by light and to the relative motion of the observer with respect to the physical system obliging the beam
to bend and come back to the observer. After having introduced the foundations of the effect, we have studied in full details a simple example to focus on the origin of the time delay in the inertial frame of the moving observer, and we have made a comparison with the case of the rotating observer. We have proved that the Sagnac effect
is not peculiar to  rotations and accelerated motion, rather it originates from the closure of the two space paths as seen in the frame co-moving with the emitter/receiver, and from relative motion between the emitter/receiver and the mirrors (or physical device); indeed, its foundations are related to the relativity of simultaneity.
On the practical side one may think of interesting applications based on the use of ring lasers as linear accelerometers.

\section*{Acknowledgments}

The authors are indebted to one of the referees for suggesting the example in section \ref{example}.

\end{document}